\documentclass{ws-p10x7}

\def\Journal#1#2#3#4{{#1} {\bf #2}, #3 (#4)}

\def\NPB{{\em Nucl. Phys.} B}

\def\PRL{\em Phys. Rev. Lett.}
\def\PRD{{\em Phys. Rev.} D}

\begin{document}

\title{Weak matrix elements and K-meson physics}

\author{Massimo Testa}
\address{Dipartimento di Fisica, Universita' di 
Roma ``La Sapienza'',\\
 P.le A.Moro 2, 00185 Roma,Italy \\ 
INFN-Sezione di Roma, Italy \\ 
E-mail: massimo.testa@roma1.infn.it}

\twocolumn[\maketitle\abstract{An overview is presented about old and 
recent methods to compute the $K\rightarrow \pi \pi$ decay amplitude.}]

\section{Introduction}
Kaon Physics is a very complicated blend of Ultraviolet and Infrared effects
which still defies complete physical understanding.

The problem consists in the large enhancement ($\approx 20$) of 
the $\Delta I={1 \over 2}$ amplitude with respect to the $\Delta I={3 \over 2}$
one.

Being a process involving hadrons, $K$-decay must be treated 
non-perturbatively, so that lattice discretization is the ideal tool 
to deal with this problem. In fact lattice regularization is the ${\it only}$ 
convergent (as $a\rightarrow 0$) approximation scheme to QCD.

Due to the difficulties of putting the Standard Model on the lattice, 
one can use weak interaction perturbation theory, which, together with 
Asymptotic Freedom of Strong Interactions, allows the 
definition of 
an effective low energy actions for non-leptonic decays:
\begin{eqnarray}
& & {\cal H}_{\Delta S=1}^{eff}=\\
& & =\lambda _u{{G_F} \over {\sqrt 2}}\left[ 
	{C_+(\mu )O^{(+)}(\mu )+C_-(\mu )O^{(-)}(\mu )} \right]\nonumber
\end{eqnarray}¥
where $\lambda _u=V_{ud}V_{us}^*$ and:
\begin{eqnarray}
& & O^{(\pm )}=\\
& & =\left[ {(\bar s\gamma _\mu ^Ld)(\bar u\gamma _\mu 
	^Lu)\pm (\bar s\gamma _\mu ^Lu)(\bar u\gamma _\mu ^Ld)} 
	\right]-\left[ {u\leftrightarrow c} \right]\nonumber
\end{eqnarray}¥
$O^{(-)}$ is a pure $I={1 \over 2}$, while $O^{(+)}$ is a mixture of 
$I={1 \over 2}$ and $I={3 \over 2}$.

The $O^{(\pm )}$\'{}s transform as $(8,1) \oplus (1,8)$ and 
$(27,1) \oplus (1,27)$ under the 
$SU(3)\otimes SU(3)$ chiral group and some discrete symmetries.
The coefficients $C_\pm (\mu )$ reliably computed in Perturbation Theory,
show a slight octet enhancement\cite{uno}:
\begin{equation}
	\left| {{{C_-(\mu \approx 2\;Gev)} \over {C_+(\mu \approx 
	2\;Gev)}}} \right|\approx 2
\end{equation}¥
The rest of the enhancement ($\approx 10$) should, then, be provided by the matrix 
elements of $O^{(\pm )}$ and is a non perturbative, infrared effect.

The difficulty with Lattice regularization lies 
in the fact that naive discretization of Dirac fermions entails a 
multiplication of low energy degrees of freedom (Doublers) whose 
elimination complicates the scheme.

There are, essentially, two possibilities:
\begin{itemize}
\item Wilson Fermions\cite{wilson}

A term is added to the Lagrangian, breaking explicitly the chiral 
symmetry, which can be restored, as $a\to 0$, by the inclusion of 
appropriate counterterms. This formulation is ultra-local 
(at the lagrangian level only near 
neighbors interactions are involved) and it is very 
convenient for numerical purposes.

\item Ginsparg-Wilson Fermions\cite{G-W}$^{,}$\cite{nn}$^{,}$\cite{n1}$^{,}$\cite{luscher}$^{,}$\cite{hasenfratz}

This discretization is much more respectful of the chiral properties 
of the (continuum) QCD lagrangian, at the expense of being non local 
at the lattice level, which makes it, at the moment, numerically very 
demanding.
\end{itemize}
My remarks on renormalization will, therefore, be addressed to Wilson fermions
formulation.

The difficulty of the problem consists, first of all, in giving the 
correct definition of the operators $O^{(\pm )}$.

In order to construct finite composite operator of dimension $6$, 
$\tilde O_6(\mu )$, we must mix the original bare operator, 
$O_6(a)$, with bare operators of equal 
($O_6^{(i)}(a)$) or smaller ($O_3(a)$) dimension, in general with different 
naive chiralities\cite{unoa}.

A general non perturbative technique to to construct composite operators
is based on the systematic exploitation of  Chiral Ward Identities
 \cite{tre}.

It turns out that, in order to minimize the renormalization procedure, 
the best strategy is to compute $K\to \pi \pi $ in the world in which 
$m_K=2m_\pi $ or $m_K=m_\pi $, with pions at rest (see 
section(\ref{infinite}))
and then extrapolate to the real world through chiral perturbation 
theory\cite{otto}$^{,}$\cite{sette}$^{,}$\cite{nove}. In these cases the ultraviolet subtractions 
are limited to an overall renormalization which could be determined 
non perturbatively\cite{cinque}.

\section{Infrared Problems}

Approaches requiring the construction of an asymptotic two pion 
state face the problems due to the fact that the theory is
defined, through the functional integral, in the euclidean region. In 
the next two subsections we will briefly discuss the nature of the 
problem and possible proposals to solve it.

\subsection{Infinite volume}\label{infinite}

In order to compute the $K\to \pi \pi $ width we have to evaluate the matrix 
element ${}_{(out)}\left\langle {\pi (\underline p)\pi (-\underline p)} 
	\right|{\cal H}_W\left| K \right\rangle$
with two interacting hadrons in the final state.
This is not easy to do in the euclidean region\cite{sei}.
It can be shown  that:
\begin{eqnarray}
	& &\left\langle {\varphi _{\underline p}(t_1)\varphi 
	_{-\underline p}(t_2){\cal H}_W(0)K(t_K)} \right\rangle \mathop \approx 
	\limits_{\scriptstyle {t_K\to -\infty }\hfill\atop
	  \scriptstyle {t_1>>t_2>>0}\hfill} \nonumber \\
	& &\approx e^{m_Kt_K-E_{\underline p}t_1-E_pt_2}\sqrt {{{Z_K} \over 
	{2m_K}}}{{Z_\pi } \over {2m_\pi }} \times \nonumber\\
	& & \left\{ {\left[ 
	{{{{}_{(out)}\left\langle {\underline p,-\underline p} 
	\right|{\cal H}_W\left| K \right\rangle +{}_{(in)}\left\langle {\underline 
	p,-\underline p} \right|{\cal H}_W\left| K \right\rangle } \over 2}} 
	\right]} \right.+ \nonumber \\
& &\left. {+P_{\underline q}(t_2)} \right. \label{asy1}
\end{eqnarray}¥
where:
\begin{eqnarray}
& & P_{\underline q}(t_2)=-{\cal P}\sum\limits_n {\exp 
	[-(E_n-2E_{\underline q})t_2](2\pi )^3\delta ^3(\underline 
	P_n)}\times \nonumber \\
& & \times N_n{{[{\cal M}(\underline q,-\underline q;n)]^*\left\langle {n,out} 
	\right|{\cal H}_W(0)\left| K \right\rangle } \over {E_n(E_n-2E_{\underline 
	q})}}   \label{asy2}
\end{eqnarray}¥
In eq.(\ref{asy2}), ${\cal M}_{\mathop {\pi \pi }\limits_{2m_\pi }\to \mathop {\pi \pi 
}\limits_{2E_\pi }}^{(o.s.)}$ denotes an off-shell extrapolation of the 
$\mathop {\pi \pi }\limits_{2m_\pi }\to \mathop {\pi \pi 
}\limits_{2E_\pi }$ scattering amplitude, which represents the euclidean 
version of the final state interaction.

The problem with eqs.(\ref{asy1}),(\ref{asy2}) is that, for large, 
positive $t_{2}$:
\begin{equation}
	P_{\underline q}(t_2)\approx e^{2(E_{\underline q}-m_\pi )t_2}
\end{equation}¥
so that, asymptotically in $t_2$, $P_{\underline q}(t_2)$ dominates 
over the physically relevant matrix elements in eq.(\ref{asy1}).

If we choose $\underline p=0$ in eq.(\ref{asy1}), we have:
\begin{eqnarray}
	& &\left\langle {\pi _{\underline 0}(t_1)\pi _{\underline 
	0}(t_2){\cal H}_W(0)K(t_K)} \right\rangle \mathop \approx 
	\limits_{\scriptstyle {t_K\to -\infty }\hfill\atop
	  \scriptstyle {t_1>>t_2>>0}\hfill}\label{matrix}\\
	& &\approx e^{m_Kt_K-m_\pi t_1-m_\pi t_2} \sqrt {{{Z_K} \over 
	{2m_K}}}{{Z_\pi } \over {2m_\pi }} \nonumber \\
	& & \left\langle {\pi 
	(\underline 0)\pi (\underline 0)} \right|{\cal H}_W\left| K \right\rangle 
	(1+{c \over {\sqrt {t_2}}}{\cal M}_{\mathop {\pi \pi }\limits_{2m_\pi }\to 
	\mathop {\pi \pi }\limits_{2m_\pi }}^{(o.s.)})\nonumber
\end{eqnarray}¥
One sees from eq.(\ref{matrix}) that for $\pi \acute {} s$ at rest and weakly interacting it is possible to 
extract a meaningful matrix element.

\subsection{Finite volume}

Lellouch and L\"{u}scher have recently formulated a strategy
 based on the exploitation of the finiteness of volume in 
lattice simulations\cite{LLL}.
Their proposal is based on the following relation between finite and
infinite volume matrix elements:
\begin{eqnarray}
& & \left| {\left\langle {\pi \pi ,E=m_K} \right|{\cal H}_W(0)\left| 
	K \right\rangle } \right|^2=\label{LLL}\\
& & =V^2\left| {{}_V\left\langle {\pi \pi ,E} \right|{\cal H}_W(0)\left| K 
	\right\rangle _V} \right|^2\left( {{{m_K} \over {k_\pi }}} 
	\right)^3 \times \nonumber \\
& &	\times 8\pi [q\phi '(q)+k\delta '_0(k)] \nonumber
\end{eqnarray}¥
In eq.(\ref{LLL}) $\left| {\pi \pi ,E} \right\rangle _V$ denotes a 
finite volume two pion state with zero total
momentum and ``angular momentum'' and energy $E$, while $\left| K \right\rangle _V$ denotes a 
single finite volume kaon state with zero momentum. Both states are normalized to $1$.
$\left| {\pi \pi ,E} \right\rangle $ and $\left| K \right\rangle $
denote the corresponding infinite volume states 
covariantly normalized according to the usual convention which, for 
single particle states reads as:
\begin{equation}
	\left\langle {{\underline p}} \mathrel{\left | {\vphantom 
	{{\underline p} {\underline q}}} \right. \kern-\nulldelimiterspace} 
	{{\underline q}} \right\rangle =(2\pi )^3 2 \omega_{\underline p} \delta ^{(3)}(\underline 
	p-\underline q)
\end{equation}¥

In a finite volume the allowed values, $k$,
of the `radial' relative momentum of a zero total momentum s-wave
two particle state obey the 
relation\cite{vol}:
\begin{equation}
	n\pi -\delta _0(k)=\phi (q) \label{phase}
\end{equation}¥
where $\delta _0(k)$ is the s-wave phase-shift, $q\equiv {{kL} \over 
{2\pi }}$, $k$ is related to the center of mass energy, $E$ as:
\begin{equation}
	E=2\sqrt {m_\pi ^2+k^2}
\end{equation}¥
and:
\begin{eqnarray}
	& &\tan \phi (q)=-{{\pi ^{{\raise3pt\hbox{$3$} 
	\!\mathord{\left/ {\vphantom {3 
	2}}\right.\kern-\nulldelimiterspace}\!\lower3pt\hbox{$2$}}}q} \over 
	{Z_{00}(1;q^2)}}\\
	& & Z_{00}(s;q^2)={1 \over {\sqrt {4\pi }}}\sum\limits_{\underline 
	n\in Z^3} {(\underline n^2-q^2)^{-s}}\label{zeta}
\end{eqnarray}¥
Eqs.(\ref{phase})-(\ref{zeta}) completely define the quantities appearing 
in eq.(\ref{LLL}).

I will now present a different approach\cite{LMST} to the relation between finite and infinite 
volume matrix elements, which may lead to a better understanding of the 
nature of eq.(\ref{LLL}). The argument goes as follows.

In order to relate the states at finite volume with those at infinite volume we 
take the two-point Green function of a scalar operator 
$\sigma (x)$, $\int\limits_V {d^3x\left\langle {\sigma (\underline 
	x,t)\sigma (0)} \right\rangle }$, and consider its behavior as the 
	space volume $V$ becomes large. We have:
\begin{eqnarray}
	& &\int\limits_V {d^3x\left\langle {\sigma (\underline 
	x,t)\sigma (0)} \right\rangle 
	_V}\mathrel{\mathop{\kern0pt\longrightarrow}\limits_{V\to \infty 
	}} \label{uno} \\
	& & {{(2\pi )^3} \over {2(2\pi )^6}}\int {{{d\underline p_1} \over 
	{2\omega _1}}{{d\underline p_2} \over {2\omega _2}}\delta (\underline 
	p_1+\underline p_2)e^{-(\omega _1+\omega _2)t}} \times \nonumber \\
& & \times {\left| {\left\langle 0 
	\right|\sigma (0)\left| {\underline p_1,\underline p_2} \right\rangle 
	} \right|^2}=\nonumber\\
	& &={1 \over {2(2\pi )^3}}\int {dE}e^{-Et}\left| {\left\langle 0 
	\right|\sigma (0)\left| {\pi \pi ,E} \right\rangle } 
	\right|^2 \times \nonumber\\
	& & \times \int 
	{{{d\underline p_1} \over {2\omega _1}}{{d\underline p_2} \over 
	{2\omega _2}}}\delta (\underline p_1+\underline p_2)\delta (E-\omega 
	_1-\omega _2)=\nonumber\\
	& &={\pi  \over {2(2\pi )^3}}\int {{{dE} \over E}}e^{-Et}\left| 
	{\left\langle 0 \right|\sigma (0)\left| E \right\rangle } 
	\right|^2k(E) \nonumber
\end{eqnarray}¥
where:
\begin{equation}
	k(E)=\sqrt {{{E^2} \over 4}-m_\pi ^2}
\end{equation}
On the other hand:
\begin{eqnarray}
	& &\int\limits_V {d^3x\left\langle {\sigma (\underline 
	x,t)\sigma (0)} \right\rangle }= \label{due} \\
& & V\sum\limits_n {\left| {\left\langle 
	0 \right|\sigma (0)\left| {\pi \pi ,n} \right\rangle _V} 
	\right|^2e^{-E_nt}}\mathrel{\mathop{\kern0pt\longrightarrow}\limits_{V\to 
	\infty }}\nonumber\\
	& &\mathrel{\mathop{\kern0pt\longrightarrow}\limits_{V\to \infty 
	}}V\int\limits_0^\infty  {dE\rho (E)\left| {\left\langle 0 
	\right|\sigma (0)\left| {\pi \pi ,E} \right\rangle _V} 
	\right|^2e^{-Et}} \nonumber
\end{eqnarray}¥
where $\left| {\pi \pi ,n} \right\rangle _V$ denote the finite volume 
two pion states classified according to the quantum number $n$ 
defined in 
eq.(\ref{phase}) and:
\begin{equation}
	\rho (E)\equiv {{\Delta n} \over {\Delta E}}={{q\phi '(q)+k\delta 
	'_0(k)} \over {4 \pi k^2}}E \label{density}
\end{equation}¥
denotes the density of states of energy $E$.

Comparing eqs.(\ref{uno}) and (\ref{due}), we get the correspondence:
\begin{equation}
	\left| {\pi \pi ,E} \right\rangle \Leftrightarrow 4\pi \sqrt 
	{{{VE\rho (E)} \over {k(E)}}}\left| {\pi \pi ,E} \right\rangle _V
\label{unoone}
\end{equation}¥
In a similar way it is easy to show:
\begin{equation}
	\left| {\underline p=0} \right\rangle \Leftrightarrow \sqrt 
	{2mV}\left| {\underline p=0} \right\rangle _V
	\label{duetwo}
\end{equation}¥
From eqs.({\ref{unoone}) and ({\ref{duetwo}) we get:
\begin{eqnarray}
& & \left| {\left\langle {\pi \pi ,E=m_K} \right|{\cal H}_W(0)\left| 
	K \right\rangle } \right|^2=\label{LLL1}\\
& & =32 \pi^{2} V^2 {{\rho(m_{K}) m_{K}^{2}}\over k_{\pi}} 
\left| {{}_V\left\langle {\pi \pi ,E} \right|{\cal H}_W(0)\left| K 
\right\rangle _V} \right|^2 \nonumber
\end{eqnarray}¥
where:
\begin{equation}
k_{\pi} \equiv \sqrt {{{m_{K}^2} \over 4}-m_\pi ^2} \nonumber
\end{equation}¥
Using the expression of $\rho (E)$ given by eq.(\ref{density}), 
eq.(\ref{LLL1}) looks the same as eq.(\ref{LLL}). There is, however an 
important difference. In fact the derivation\cite {LLL} of eq.(\ref{LLL}) requires to 
work at a fixed volume $V$ and at a fixed
value of $n$, defined in eq.(\ref{phase})\footnote{in fact $n < 8$}.
Eq.(\ref{LLL1}), on the contrary, is 
valid at fixed energy $E$, asymptotically in $V$, so that, while we 
let $V \rightarrow \infty$, we must allow simultaneously $n \rightarrow 
\infty$. A possible relation between the two approaches will be 
discussed in a forthcoming paper\cite{LMST}.

The strategy proposed by Lellouch and L\"{u}scher\cite{LLL} consists in
tuning the volume $V$ so that the first excited two-pion state 
($n=1$) is degenerate in energy with the kaon state 
($L\approx 5\div 6\;Fm$) and compute the finite volume Green\'{}s function:
\begin{eqnarray}
& &\int\limits_V {d^3xd^3y\left\langle {\sigma (\underline 
x,t){\cal H}_W(0)K(\underline y,t')} \right\rangle }_V\mathop \approx 
\limits_{t'\to -\infty } \nonumber \\
& & =e^{m_Kt'}{}_V\left\langle K \right|K(0)\left| 0 \right\rangle 
V^2\times  \nonumber \\
& & \times\sum\limits_n {\left\langle 0 \right|\sigma (0)\left| {\pi \pi ,n} 
\right\rangle _V}\times  \nonumber \\
& &\times {}_V\left\langle {\pi \pi ,n} \right|{\cal H}_W(0)\left| K \right\rangle 
_Ve^{-E_nt}= \nonumber \\
& & = e^{m_Kt'}{}_V\left\langle K \right|K(0)\left| 0 \right\rangle 
V^2\times\nonumber \\
& & \times \sum\limits_n {\left| {\left\langle 0 \right|\sigma (0)\left| {\pi \pi ,n} \right\rangle _V} \right|}\times  \nonumber \\
& & \times \left| {{}_V\left\langle {\pi \pi ,n} \right|{\cal H}_W(0)\left| K \right\rangle _V} \right|e^{-E_nt}
\label{cancel}
\end{eqnarray}
The last equality in eq.(\ref{cancel}) is justified by the 
cancellation of the final state interactions phases in
 ${\left\langle 0 \right|\sigma (0)\left| {\pi \pi ,n} \right\rangle _V}$ and 
 ${{}_V\left\langle {\pi \pi ,n} \right|{\cal H}_W(0)\left| K \right\rangle _V}$.

Then, from
\begin{eqnarray}
& & \int\limits_V {d^3x\left\langle {\sigma (\underline x,t)\sigma (0)} 
	\right\rangle }=\\
& & V\sum\limits_n {\left| {\left\langle 0 \right|\sigma 
	(0)\left| {\pi \pi ,n} \right\rangle _V} \right|^2e^{-E_nt}}\nonumber
\end{eqnarray}
we compute $\left| {\left\langle 0 \right|\sigma (0)\left| {\pi \pi ,1} 
\right\rangle _V} \right|$ and, finally, 
$\left| {{}_V\left\langle {\pi \pi ,1} \right|{\cal H}_W(0)\left| K 
\right\rangle _V} \right|$.

In the case of a $\Delta I={1 \over 2}$ transition we face
a further complication\cite{sette}$^{,}$\cite{nove}: independently of the 
chosen procedure, a subtraction has to be performed, due to the fact that the relevant 
correlator $ \left\langle {\sigma (t) {\cal H}_W(0) K(t')} 
\right\rangle $ is dominated, for large $t$, by the vacuum insertion 
between $\sigma (t)$ and ${\cal H}_W(0)K(t')$. As a consequence, the relevant 
physical information about the $K$ decay is contained in the 
connected correlator:
\begin{eqnarray}
& & \left\langle {\sigma (t){\cal H}_W(0)K(t')} \right\rangle 
	_{conn} \equiv \\
& & \equiv  \left\langle {\sigma (t) {\cal H}_W(0) K(t')} \right\rangle -
	\left\langle {\sigma (0)} \right\rangle \left\langle 
	{{\cal H}_W(0)K(t')} \right\rangle \nonumber
\end{eqnarray}¥

\end{document}